\documentclass[twocolumn,floatfix]{revtex4-1}

\usepackage{graphicx}
\usepackage{xcolor}
\usepackage{amsmath}
\usepackage{amssymb}

\newcommand{\be}{\begin{equation}}
\newcommand{\ee}{\end{equation}}
\newcommand{\ba}{\begin{array}{l}}
\newcommand{\ea}{\end{array}}

\newcommand{\banonum}{\begin{eqnarray*}}
\newcommand{\eanonum}{\end{eqnarray*}}
\newcommand{\baa}{\begin{eqnarray}}
\newcommand{\eaa}{\end{eqnarray}}
\newcommand{\bfr}{\begin{flushright}}
\newcommand{\efr}{\end{flushright}}
\newcommand{\bfl}{\begin{flushleft}}
\newcommand{\efl}{\end{flushleft}}
   \def\sH{{\mathfrak H}}   
      
   \def\sN{{\mathfrak N}}

      \def\dC{{\mathbb C}}

   \def\dN{{\mathbb N}}   
      \def\dR{{\mathbb R}}

\def\cA{{\mathcal A}}      
      
   \def\cH{{\mathcal H}}

\newcommand {\wt}{\widetilde}
\newcommand {\wh}{\widehat}
\DeclareMathOperator{\dom}{dom}

\newtheorem{theorem}{Theorem}

\newtheorem{corollary}[theorem]{Corollary}
\newtheorem{definition}[theorem]{Definition}
\newtheorem{remark}[theorem]{Remark}

\newenvironment{proof}[1][Proof]{\begin{trivlist}
\item[\hskip \labelsep {\bfseries #1}]}{\end{trivlist}}

\begin{document}

\title{Transparent boundary conditions for the stationary Schr\"{o}dinger equation via Weyl-Titchmarsh theory}
\author{V.A.~Derkach$^{1}$, C.~Trunk$^{1}$, J.R.~Yusupov$^{2}$, D.U.~Matrasulov$^{3}$}
\affiliation{$^1$Technical University of Ilmenau, 25 Weimarer Str., 100565, D-98693, Ilmenau, Germany\\
$^2$Kimyo International University in Tashkent, 156 Usman Nasyr Str., 100121, Tashkent, Uzbekistan\\
$^3$Turin Polytechnic University in Tashkent, 17 Niyazov Str., 100095, Tashkent, Uzbekistan}

\thanks{
  \includegraphics[height=7.0mm]{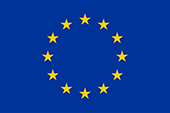} This paper is supported by European Union’s Horizon 2020 research and innovation programme under the Marie Sklodowska-Curie grant agreement ID: 873071, project SOMPATY (Spectral Optimization: From Mathematics to Physics and Advanced Technology)
}

\begin{abstract}
We propose a general approach for deriving transparent boundary conditions for the stationary Schr\"odinger equation with arbitrary potential. It is proven that the transparent boundary conditions can be written in terms of the Weyl-Titchmarsh coefficients. 
As examples for the application of the proposed approach, two special cases for the stationary Schr\"odinger equation with the harmonic potential and the P\"oschl-Teller potential are considered.
\end{abstract}

\maketitle

\section{Introduction}

Controlling the quantum particle dynamics in low-dimensional domains is of practical importance for tuning of functional properties of advanced materials and device optimization in emerging quantum technologies. One of the attractive problems in this topic is the absence of backscattering in the border of a given domain during the transmission of a quantum particle through the domain's boundary. 

By solving such a task, one can minimize losses in the transport of quasiparticles, which serve as carriers of signals, energy, information, and other physical characteristics in various quantum functional materials. 
Traditional way for modelling of quantum transport in quantum mechanics is using a scattering (S-) matrix-based approach, where the elements of the S-matrix describe transmission and reflection probabilities. However, to use such an approach, one needs to have an exact or asymptotic solution of the Schr\"odinger equation. 

An alternative approach based on description of reflectionless transmission of particles and waves through the given domain was developed in mathematical physics during the past two decades.  Within this approach, one imposes so-called transparent boundary conditions for a given evolution equation that ensures reflectionless transmission of wave or particle through the boundary of the domain. 
Since the appearance of the pioneering papers \cite{Engquist1,Engquist2}, the concept of transparent boundary conditions (TBC) was applied to various linear \cite{Gander,Antoine1, Arnold1998,Ehrhardt1999,Ehrhardt2001,Hammer2014,Wu,Schwendt} and nonlinear \cite{Han0,Antoine,Matthias2008,Zhang, Han,Li, Zheng07} evolution equations. Within this approach, it is required that the solution of a given evolution equation on a finite interval of the computational domain matches the solution on the whole space restricted to that interval. This leads to the boundary conditions, whose explicit form is quite complicated. Moreover, solving the wave equation numerically with such boundary conditions requires the use of highly effective and stable discretization schemes. However, in some cases, the transparent boundary conditions can be replaced by physically equivalent ones, which have a much simpler form with clearly expressed physical meaning. This was shown recently, e.g., for quantum graphs in \cite{Jambul}. Earlier, the concept of transparent boundary conditions has been extended to time-independent Schr\"odinger equation with general (linear and nonlinear) potential, by considering spectral problems \cite{Klein2011}. Within such a concept, the ``transparency" implies that the eigenvalue spectrum of the problem on the considered finite domain coincides with that in the whole space.

In this paper, we address the problem of transparent boundary conditions for the stationary Schr\"odinger equation that contains a time-independent potential. Unlike the usual approach for the derivation of transparent boundary conditions, we use the Weyl-Titchmarsh function that greatly facilitates the derivation of transparent boundary conditions. In addition, in such an approach TBC have a rather simple form and can be physically well implemented. Motivation for the study of transparent boundary conditions for the time-independent Schr\"odinger equation with interaction potentials comes from several problems of condensed matter physics and optics, where tuning of the electronic band structure and ballistics quantum transport need to be achieved. 

The paper is organized as follows.
The next section provides some preliminaries on the Weyl-Titchmarsh theory. Sections~\ref{Shymbulak} and~\ref{Astana} present derivations of transparent boundary conditions for time-independent Schr\"odinger equations with a discrete spectrum potential and in the general case, respectively. In Section~\ref{Examples} we apply transparent boundary conditions to Schr\"odinger equation with harmonic potential and P\"oschl-Teller potential. Finally,  there is an appendix on boundary triplets, Weyl functions and the so-called coupling method. This provides the machinery to extend the characterization of transparent boundary conditions given in Sections~\ref{Shymbulak} and~\ref{Astana} to a wider class of differential operators, including operators of higher order and partial differential operators.

\section{Preliminaries. Sturm-Liouville   operators and Weyl-Titchmarsh coefficients}\label{Almaty}

Let us first recall the notion of the Weyl-Titchmarsh coefficient for a Sturm-Liouville  differential operator. For the half-lines $I_-=(-\infty,a_-]\subset\dR$ and $I_+=[a_+,+\infty)\subset\dR$
we denote by $A_\pm$ the \emph{minimal operator} associated in $L^{2}(I_\pm)$ with the differential expression
\begin{equation}\label{eq:100.13}
\mathcal{A}_\pm=-\frac{d^2}{dx^2}+V_\pm
\end{equation}
with real-valued coefficient  $V_\pm\in L_{\mathrm{loc}}^{1}(I_\pm)$, see~\cite{Naj69}.
The operator $A_\pm$ is closed and symmetric, but is not self-adjoint in $\sH_\pm:=L^{2}(I_{\pm})$.
Hence $A_\pm\subset A_{\pm}^*$ and the domain $D(A_{\pm}^*)$ of the \emph{maximal operator} $A_{\pm}^*$ consists of functions $f\in L^{2}(I_\pm)$ such that $f$ and $f'$ are locally absolutely continuous on $I_\pm$ and $\cA_\pm f\in L^{2}(I_\pm)$. Let
$$
\wh\rho(A_\pm):=\{\lambda\in\dC: \text{ range of }
A_{\pm}-\lambda I_{\pm}\text{ is closed in }\sH_\pm\} 
$$
be the set of regular type points of $A_\pm$. 
The subspace 
\begin{equation}\label{eq:def_sub}
\sN_{\lambda}(A_{\pm})=\ker(A_\pm^*-\lambda I_\pm),\quad \lambda\in\wh\rho(A_\pm),
\end{equation}
is called the defect subspace of $A_{\pm}$.
Here $\ker (A_\pm^*-\lambda I_\pm)$ is the kernel,
i.e.\ the set of all vectors from the domain $D(A_{\pm}^*)$ which are mapped by $A_\pm^*-\lambda I_\pm$ to zero. Note, that 
the functions $\lambda \mapsto \dim\sN_{\lambda}(A_\pm)$ is constant for $\lambda \in \dC_+$
and constant in $\dC_-$,
where $\dC_+$ and $\dC_-$ stands for the open upper (resp.\ lower) half-plane. The numbers
\begin{equation}\label{party}
n_+(A_{\pm}):=\dim\sN_{i}(A_\pm),\ 
n_-(A_{\pm}):=\dim\sN_{-i}(A_\pm)
\end{equation}
are  called the defect numbers of $A_{\pm}$.
It is known, see, e.g., \cite{Naj69}, that 
$$
n_{+}(A_+)=n_{-}(A_+) \quad \text{and} \quad
n_{+}(A_-)=n_{-}(A_-)
$$
and, due to Weyl's Alternative,
the above numbers take only the values $1$ or $2$.
 The expression $\mathcal{A}_\pm$ is said to be in the limit point case at $\pm\infty$, if $n_\pm(A_\pm)=1$.
 In this case, the domain of the minimal operators 
 $A_\pm$ consists of all functions   $\psi \in L^{2}(I_\pm)$ such that $\psi$ and $\psi'$ are locally absolutely continuous on $I_\pm$, $\cA_\pm \psi  \in L^{2}(I_\pm)$  and $\psi(a_\pm) = \psi^\prime(a_\pm)=0$. 
Let ${\lambda} \in \mathbb{C}\setminus\mathbb{R}$ and denote by ${s}_{\pm}({\lambda},\cdot)$ and ${c}_{\pm}({\lambda},\cdot)$ the solutions on $I_\pm$ of the equations
\begin{equation} \label{eq:2.13a}
  \cA_\pm(f)={\lambda} f,
\end{equation}
satisfying the boundary conditions
 \[
{c}_{\pm}({\lambda},a_{\pm}) = s_{\pm}'({\lambda},a_{\pm}) = 1, \quad c_{\pm}'({\lambda},a_{\pm}) ={s}_{\pm}({\lambda},a_{\pm}) = 0 .
 \]
If  $\mathcal{A}_\pm$ is in the limit point case at $\pm\infty$, then, by~\cite{Tit62}, neither ${s}_\pm({\lambda},\cdot)$ nor ${c}_\pm({\lambda},\cdot)$ belong to  $L^2(I_\pm)$, however there exists a coefficient $m_\pm({\lambda})$ such that the  solution
\begin{equation} \label{eTWeyl}
\psi_\pm({\lambda},x) = {c}_\pm({\lambda},x) \pm m_\pm({\lambda}) {s}_\pm({\lambda},x), \quad x
\in I_\pm,
\end{equation}
of \eqref{eq:2.13a} belong to $L^2(I_\pm)$ and also to $\sN_{\lambda}(A_\pm)$ as $\dim\sN_{\lambda}(A_\pm)=1$ for all non-real $\lambda$. Hence
$$
\sN_{\lambda}(A_\pm) = \text{span}
\{ \psi_\pm({\lambda},\cdot) \}.
$$
The function $m_\pm$ is called
the Weyl-Titchmarsh coefficient of $A_\pm$.
The Weyl-Titchmarsh coefficients $m_\pm$ belong to the Herglotz-Nevanlinna class of functions $m$ holomorphic in $\dC_+\cup\dC_-$ such that
\[
m(\overline{\lambda})=\overline{m(\lambda)}\quad\text{and}\quad \text{Im }m(\lambda)\ge 0\quad\text{for all $\lambda\in\dC_+$.}
\]

Notice that $m_\pm$ admit an extension to a meromorphic function on 
$\wh\rho(A_\pm)$.
Let us mention one important formula for the Weyl-Titchmarsh coefficient $m_\pm$. Let  $\Gamma_0^\pm $ and $\Gamma_1^\pm$ be trace functionals defined on
$ D(A_{\pm}^*)$ by
\begin{equation} \label{eq:BTforSL}
\Gamma_0^\pm f= f(a_\pm),\quad \Gamma_1^\pm f= \pm f'(a_\pm),\quad f\in D(A_\pm^*).
\end{equation}
Then for every  ${\lambda} \in \mathbb{C}\setminus\mathbb{R}$  we have
\begin{equation} \label{eq:WeylCoef}
\Gamma_0^\pm \psi_\pm({\lambda},\cdot)= 1,\quad \Gamma_1^\pm \psi_\pm({\lambda},\cdot)=  m_\pm({\lambda})
\end{equation}
and hence
\begin{equation}\label{eq:Weyl_Func}
m_\pm({\lambda})=\frac{\Gamma_1^\pm \psi_\pm({\lambda},\cdot)}{\Gamma_0^\pm \psi_\pm({\lambda},\cdot)}.
\end{equation}
Another important issue is the second Green formula which holds
for all $f, g\in D(A_\pm^*)$
\begin{equation}\label{Greendef}
(A_\pm^*f, g)_{\sH_\pm}-(f, A_\pm^*g)_{\sH_\pm} = \Gamma_1^\pm f\,\overline{\Gamma_0^\pm g}- \Gamma_0^\pm f\,\overline{\Gamma_1^\pm g}.
\end{equation}

This formula motivates the definition of a boundary triple for a symmetric operator which will be discussed in the Appendix.

\section{Transparent boundary conditions for discrete Schr\"{o}dinger operators}
\label{Shymbulak}

In this section we will focus on one-dimensional stationary Schr\"{o}dinger equation
\begin{equation}\label{eq::1}
\left(-\frac{d^2}{dx^2}+V\right)\psi=E\psi, \ \ \ x\in \mathbb{R},
\end{equation}
with real-valued coefficient  $V\in L_{\mathrm{loc}}^{1}(\dR)$. A pair  $(\psi, E)$ is called a solution of~\eqref{eq::1} if $E\in\dR$ and  $\psi$ and $\psi'$ are locally absolutely continuous on $\dR$ such that $\psi\in L^{2}(\dR)$ and~\eqref{eq::1} holds. Suppose that
\begin{enumerate}
  \item [(H1)] the differential expression $\mathcal{A}=-{d^2}/{dx^2}+V$ is in the limit point case at $\pm\infty$;
  \item [(H2)] the self-adjoint operator $H$ generated by the differential expression $\mathcal{A}$
in $L^2({\dR})$ has a discrete spectrum.
\end{enumerate}

Note, that by (H1), the domain of $H$ consists of all functions $\psi \in L^{2}(\dR)$ such that $\psi$ and $\psi'$ are locally absolutely continuous and $\cA \psi  \in L^{2}(\dR)$ .

Alongside with the eigenvalue problem~(\ref{eq::1}) associated to the differential expression $\mathcal{A}$ on the real line $\dR$, we also consider an eigenvalue problem associated with the differential expression $\mathcal{A}$ on the finite interval ${I}:=(a_-,a_+)$.

\begin{definition}\label{def:3.2}
Let $\tau_-$ and $\tau_+$ be a pair of meromorphic in $\mathbb C$ Herglotz-Nevanlinna functions.
A number $E\in\dR$ is called an eigenvalue of the problem
\begin{equation}\label{eq:4.EV}
  (\cA- E I)\varphi=0\quad \text{on}\quad I
\end{equation}
with energy dependent boundary conditions
\begin{align}\label{eq:TBC1}
      \varphi'(E,a_-)+\tau_-(E)\varphi(E,a_-)=0,\\
      \varphi'(E,a_+)-\tau_+(E)\varphi(E,a_+)=0,\label{eq:TBC2}
\end{align}
if there exists
a non trivial function $\varphi\in L^{2}(I)$
such that  $\varphi$ and $\varphi'$ are locally absolutely continuous on $I$, which is a solution
of equation~\eqref{eq:4.EV} satisfying~\eqref{eq:TBC1}--\eqref{eq:TBC2}.
If $E$ is a pole of $\tau_-$ $($resp., $\tau_+)$ then equation \eqref{eq:TBC1} $($resp., \eqref{eq:TBC2}$)$ is understood as
\[
\varphi(E,a_-)=0\quad (\text{resp.,}\,\, \varphi(E,a_+)=0).
\]
The function $\varphi$ is called the eigenfunction of the problem~\eqref{eq:4.EV}--\eqref{eq:TBC2} corresponding to the eigenvalue $E$.
\end{definition}

\begin{definition}\label{def:TBC}
Let the differential expression $\mathcal{A}$ satisfy the hypotheses $(H1)$ and $(H2)$, let $I=(a_-,a_+)$ and $H$ be the self-adjoint operator generated by the differential expression $\mathcal{A}$ in $L^2({\dR})$. Let $\tau_-$ and $\tau_+$ be Herglotz-Nevanlinna functions.  The energy dependent boundary conditions \eqref{eq:TBC1}, \eqref{eq:TBC2} are called transparent for $\mathcal{A}$ on $I$, if:
\begin{enumerate}
    \item[(1)]  for every  solution $(\psi, E)$ for~\eqref{eq::1} the restriction $\varphi$ of $\psi$ to $I$ is an eigenfunction for the problem~\eqref{eq:4.EV}--\eqref{eq:TBC2};
    \item[(2)]  for every eigenfunction $\varphi$ for the problem~\eqref{eq:4.EV}--\eqref{eq:TBC2} there exists a solution $(\psi, E)$ for~\eqref{eq::1} such that $\varphi$ is the restriction of $\psi$ to $I$;
    \item[(3)] the spectrum of the problem~\eqref{eq:4.EV}--\eqref{eq:TBC2} 
    $($i.e.\ the collection of all eigenvalues corresponding to eigenfunctions of \eqref{eq:4.EV}--\eqref{eq:TBC2}$)$
    coincides with the spectrum of $H$.
\end{enumerate}
\end{definition}
Note that, by Definition~\ref{def:3.2}, item  (3) is a consequence of (1) and (2).

In the following theorem we present transparent boundary conditions for the differential expression $\mathcal{A}$ satisfying the hypotheses $(H1)$ and $(H2)$.
\begin{theorem}\label{thm:3.1}
Let the differential expression $\mathcal{A}=-\frac{d^2}{dx^2}+V$ satisfy (H1) and (H2), let $(a_-,a_+)$ be a finite interval in $\mathbb{R}$, and let $m_-$ and $m_+$ be the Weyl-Titchmarsh coefficients  for $A_\pm$, cf.\ \eqref{eTWeyl}. 
Then transparent boundary conditions for $\mathcal{A}$ on the interval 
 $I=(a_-,a_+)$ can be chosen as
\begin{align}\label{eq:TBC1.1}
     \varphi'(E,a_-)+m_-(E)\varphi(E,a_-)=0,\\
     \varphi'(E,a_+)-m_+(E)\varphi(E,a_+)=0.\label{eq:TBC2.1}
\end{align}
\end{theorem}

\begin{proof}
\textbf{1. }
By (H1) the defect numbers of $A_\pm$ are equal to $1$, cf.\ Section~\ref{Almaty}. By (H2), the sets $\wh\rho(A_\pm)$ of regular type points of $A_\pm$  coincide with $\dC$,  see~\cite{Naj69}, and, hence, the defect subspaces $\sN_\lambda(A_\pm)$ are well defined by~\eqref{eq:def_sub} for every $\lambda\in\dC$. 
\bigskip

\noindent
\textbf{2. }
Let  $(\psi, E)$ be a solution for~\eqref{eq::1} with $\psi\not\equiv 0$.  
Decompose $\psi(E,\cdot)$ as follows
\[
\psi(E,\cdot)=\begin{pmatrix}
\psi_-(E,\cdot) & \varphi(E,\cdot) & \psi_+(E,\cdot)
\end{pmatrix},
\]
where $\psi_\pm(E,\cdot)\in L^2(I_\pm)$, $\varphi(E,\cdot) \in L^2(I)$. Then
\begin{equation}\label{eq:Def+-}
\mathcal{A}\psi_-=E\psi_-,\quad
\mathcal{A}\varphi=E\varphi,\quad
\mathcal{A}\psi_+=E\psi_+,\quad
\end{equation}
and since $\psi$ and $\psi'$ are continuous, we have
 \begin{align}\label{eq:BC1}
        \varphi(E,a_-)=\psi_-(E,a_-),\\
        \label{eq:BC2}
        \varphi'(E,a_-)=\psi_-'(E,a_-),\\
        \label{eq:BC3}
        \varphi(E,a_+)=\psi_+(E,a_+),\\
        \label{eq:BC4}
        \varphi'(E,a_+)=\psi_+'(E,a_+).
\end{align}
It follows from \eqref{eq:Def+-} that $\psi_\pm(E,\cdot)\in \mathfrak{N}_E(A_\pm)$ 
and hence, by~\eqref{eTWeyl},
\begin{align}\label{eq:Def_+}
\psi_+'(E,a_+)&=m_{+}(E)\psi_+(E,a_+),\\
\psi_-'(E,a_-)&=-m_{-}(E)\psi_-(E,a_-).\label{eq:Def_-}
\end{align}
Therefore, it follows from \eqref{eq:Def+-} and \eqref{eq:BC1}-\eqref{eq:Def_-} that $\varphi(E,\cdot)$ is a solution of \eqref{eq:4.EV}, such that
\[
\begin{split}
    \varphi'(E,a_-)=\psi_-'(E,a_-)
    &=-m_{-}(E)\psi_-(E,a_-)\\
    &=-m_{-}(E)\varphi(E,a_-),
\end{split}
\]
\[
\begin{split}
    \varphi'(E,a_+)=\psi_+'(E,a_+)
    &=m_{+}(E)\psi_+(E,a_+)\\
    &=m_{+}(E)\varphi(E,a_+).
\end{split}
\]
The function $\varphi$ is not trivial since otherwise we obtain, by \eqref{eq:BC1}--\eqref{eq:BC3}, that $\psi_\pm\equiv 0$ and hence that $\psi\equiv 0$. Therefore, $E$ is an eigenvalue of the problem~\eqref{eq:4.EV}--\eqref{eq:TBC2}. 
\bigskip

\noindent
\textbf{3. } Conversely, let $E$ be an eigenvalue of the problem~\eqref{eq:4.EV}, \eqref{eq:TBC1.1}-\eqref{eq:TBC2.1} and let $\varphi(E,\cdot)$ be the corresponding eigenfunction. Consider functions
$\psi_\pm(E,\cdot)\in\sN_E(A_\pm)$ 
such that
\eqref{eq:BC1} and \eqref{eq:BC3} hold.
Then it follows from \eqref{eq:Def_+}, \eqref{eq:BC3}  and \eqref{eq:TBC2.1} that 
\[
\begin{split}
    \psi_+'(E,a_+)
    &=m_{+}(E)\psi_+(E,a_+)\\
    &=m_{+}(E)\varphi(E,a_+)=\varphi'(E,a_+).
\end{split}
\]
Similarly, \eqref{eq:Def_-}, \eqref{eq:BC1} and \eqref{eq:TBC1.1} yield
\[
\begin{split}
    \psi_-'(E,a_-)
    &=-m_{-}(E)\psi_-(E,a_-)\\
    &=-m_{-}(E)\varphi(E,a_-)=\varphi'(E,a_-).
\end{split}
\]
Therefore, the functions
\[
\psi(E,\cdot)=\begin{pmatrix}
\psi_-(E,\cdot) & \varphi(E,\cdot) & \psi_+(E,\cdot)
\end{pmatrix}
\]
and $\psi'(E,\cdot)$ are continuous on $\mathbb{R}$, $\psi(E,\cdot)\in\ker{(H-E I)}$ and $\varphi(E,\cdot)=\psi(E,\cdot)|_{I}$.
Since $\psi\not\equiv 0$,  $E$ is an eigenvalue of the operator $H$.
\end{proof}

\begin{corollary}\label{cor:4.3}
The set of eigenvalues of the problem~~\eqref{eq:4.EV}, \eqref{eq:TBC1.1}-\eqref{eq:TBC2.1} coincides with 
the set of all eigenvalues $\sigma_p(H)$ of the self-adjoint operator $H$.
\end{corollary}
\begin{remark}
\begin{itemize}
 \item[(1)] During the last three decades artificial boundary conditions for different classes of operators have been studied in many papers, see reviews~\cite{Givoli91} and \cite{Tsynkov98}. In particular, in the unpublished manuscript~\cite{EhrZhe09} transparent boundary conditions~\eqref{eq:TBC1.1}--\eqref{eq:TBC2.1} were obtained by another method.
 \item[(2)] However, for computational purposes, one has to find good approximations for exact transparent boundary conditions, see~\cite{Zhe07}, \cite{Klein2011}.
 In particular, in~\cite{Klein2011}  the so-called second-order absorbing boundary conditions  for the stationary Schr\"odinger equation were considered:
\begin{align}\label{eq:ABC1.1}
     \varphi'(E,a_-)+i\sqrt{E-V(a_-)}\varphi(E,a_-)=0,\\
     \varphi'(E,a_+)-i\sqrt{E-V(a_+)}\varphi(E,a_+)=0.\label{eq:ABC2.1}
\end{align}
In the case when $V\in C^1[a_{\pm},a_{\pm}\pm\varepsilon]$ $(\varepsilon>0)$ the Weyl-Titchmarsh coefficient $m_{\pm}(E)$ has the asymptotic expansion (can be derived from~\cite{Atk81}, see also \cite{Har84})
\[
m_{\pm}(E)=i\sqrt{E}+\frac{V(a_\pm)}{2i\sqrt{E}} +{ O}\left(\frac{1}{E}\right)\quad\text{as}\quad E\to\infty.
\]
Since
\[
i\sqrt{E-V(a_+)}=i\sqrt{E}+\frac{V(a_\pm)}{2i\sqrt{E}}+{ o}\left(\frac{1}{E}\right)\quad\text{as}\quad E\to\infty,
\]
the absorbing boundary conditions \eqref{eq:ABC1.1}--\eqref{eq:ABC2.1} can be treated as an approximation for the exact transparent boundary conditions \eqref{eq:TBC1.1}--\eqref{eq:TBC2.1}.
\end{itemize}
\end{remark}

\section{General case}
\label{Astana}

In this section, we will drop the assumption (H2) and consider the one-dimensional stationary Schr\"{o}dinger operation which satisfies (H1). Namely, for $g\in L^2(\dR)$ and $\lambda\in\dC_+\cup\dC_-$ consider the differential equation
\begin{equation}
\left(-\frac{d^2}{dx^2}+V-\lambda\right)\psi=g, \ \ \ x\in \mathbb{R},\label{eq::1.55}
\end{equation}
with a  potential $V\in L^1_{\rm loc}(\mathbb{R})$.
As before, we set $\mathcal{A}:=-{d^2}/{dx^2}+V$. Then the operator $H$ generated by the differential expression $\mathcal{A}$ with a domain $\dom H$
consisting of all functions $\psi \in L^{2}(\dR)$ such that $\psi$ and $\psi'$ are locally absolutely continuous and $\cA \psi  \in L^{2}(\dR)$
is self-adjoint in $L^2(\dR)$. Therefore, since $\lambda$ is chosen as non-real, problem~\eqref{eq::1.55} has a unique solution $\psi$ for every $g\in L^2(\dR)$.

Now, let us assume that $g\in L^2({I})$, where ${I}=[a_-,a_+]$ is a finite interval and consider  the differential equation
\begin{equation}\label{eq:Afg2_4}
  (\cA- \lambda I)\varphi=g\quad \text{on}\quad I
\end{equation}
and the corresponding equation on the whole line
\begin{equation}\label{eq:Afg.3_4}
  (\cA- \lambda I)\psi=\wt g\quad \text{on}\quad \dR,
\end{equation}
where $\wt g$ is the trivial extension of $g$ to $\dR$: $\wt g=g\oplus 0_{I_-\cup I_+}$.
A solution of~\eqref{eq:Afg.3_4} is a function 
$\psi\in \dom H$ which satisfies~\eqref{eq:Afg.3_4} in $L^2(\dR)$.
Hence, if $g=0$, a solution $\psi$ of~\eqref{eq:Afg.3_4} is an element of $\ker (H-\lambda I)$.

\begin{definition}\label{def:3.2.C}
Let $\tau_-$ and $\tau_+$ be Herglotz-Nevanlinna functions meromorphic on $\wh\rho(A_-)\cap\wh\rho(A_+)$ and let $\lambda\in\wh\rho(A_-)\cap\wh\rho(A_+)$, $g\in L^2(I)$.
A function $\varphi(\lambda,x)$ is called a solution of the problem~\eqref{eq:Afg2_4}
with energy dependent boundary conditions
\begin{align}\label{eq:TBC10.2_4}
      \varphi'(\lambda,a_-)+\tau_-(\lambda)\varphi(\lambda,a_-)=0,\\
      \varphi'(\lambda,a_+)-\tau_+(\lambda)\varphi(\lambda,a_+)=0
      \label{eq:TBC10.3_4}
\end{align}
there exists a function $\varphi$ such that $\varphi$ and $\varphi'$ are locally absolutely continuous on $I$, $\varphi\in L^{2}(I)$ and~\eqref{eq:Afg2_4}, \eqref{eq:TBC10.2_4}, \eqref{eq:TBC10.3_4} hold.
A number $\lambda\in\wh\rho(A_-)\cap\wh\rho(A_+)$ is called a regular point of the problem~\eqref{eq:Afg2_4}, \eqref{eq:TBC10.2_4}, \eqref{eq:TBC10.3_4} if for every $g\in L^2(I)$ there exists a unique solution $\varphi$ of the problem~\eqref{eq:Afg2_4}, \eqref{eq:TBC10.2_4}, \eqref{eq:TBC10.3_4}.
\end{definition}
Recall, that if $\lambda$ is a pole of $\tau_-$ or $\tau_+$ then equation \eqref{eq:TBC1} or \eqref{eq:TBC2} are understood as Dirichlet conditions
\[
\varphi(\lambda,a_-)=0\quad \text{or}\quad \varphi(\lambda,a_+)=0.
\]

\begin{definition}\label{def:4.1}
Let the differential expression $\mathcal{A}$ satisfy the hypothesis $(H1)$, let $I=(a_-,a_+)$ and $H$ be the self-adjoint operator generated by the differential expression $\mathcal{A}$ in $L^2({\dR})$.
Let $\tau_-$ and $\tau_+$ be Herglotz-Nevanlinna functions. The
energy dependent boundary conditions
\eqref{eq:TBC10.2_4} and~\eqref{eq:TBC10.3_4}
are called transparent for $\mathcal{A}$ on $I$, if:
\begin{enumerate}
    \item[(1)]  for $\lambda\in\wh\rho(A_-)\cap\wh\rho(A_+)$, $g\in L^2(I)$ and its trivial extension $\wt g$ to $\dR$,
    $$
    \wt g:=g\oplus 0_{I_-\cup I_+},
    $$
     for every solution $\psi$ for~\eqref{eq:Afg.3_4} the restriction $\varphi$ of $\psi$ to $I$ is a  solution  for the problem~\eqref{eq:Afg2_4},
    \eqref{eq:TBC10.2_4}, \eqref{eq:TBC10.3_4};
    \item[(2)]  for every solution $\varphi$  of the problem~\eqref{eq:Afg2_4},
    \eqref{eq:TBC10.2_4}, \eqref{eq:TBC10.3_4} 
 there exists a solution $\psi$ for~\eqref{eq:Afg.3_4} such that $\varphi$ is the restriction of $\psi$ to $I$;
    \item[(3)] the set of regular points of the problem~\eqref{eq:Afg2_4},
    \eqref{eq:TBC10.2_4}, \eqref{eq:TBC10.3_4} coincides with the resolvent set $\rho(H)$  of $H$.
\end{enumerate}
\end{definition}

\begin{theorem}\label{thm:3.1.Cont}
Let the differential expression $\mathcal{A}=-\frac{d^2}{dx^2}+V$ satisfy (H1), let $(a_-,a_+)$ be a finite interval in $\mathbb{R}$, and
$\lambda\in\wh\rho(A_-)\cap\wh\rho(A_+)$.
Let $m_-$ and $m_+$ be the Weyl-Titchmarsh coefficients for $A_\pm$, cf. \eqref{eTWeyl}.
Then the unique pair of 
transparent boundary conditions for the interval $I=(a_-,a_+)$ have the form
\begin{align}\label{eq:TBC10A}
     \varphi'(\lambda,a_-)+m_-(\lambda)\varphi(\lambda,a_-)=0,\\
     \varphi'(\lambda,a_+)-m_+(\lambda)\varphi(\lambda,a_+)=0,
     \label{eq:TBC10B}
\end{align}
\end{theorem}

\begin{proof}
\textbf{1. }
Let $\lambda\in\wh\rho(A_-)\cap\wh\rho(A_+)$, $g\in L^2(I)$ and let $\wt g$ be its trivial extension to $\dR$.
Let $\psi(\lambda,\cdot)$ be a solution of~\eqref{eq:Afg.3_4}. 
Decompose $\psi(\lambda,\cdot)$ as follows
\[
\psi(\lambda,\cdot)=\begin{pmatrix}
\psi_-(\lambda,\cdot) & \varphi(\lambda,\cdot) & \psi_+(\lambda,\cdot)
\end{pmatrix},
\]
where $\psi_\pm(\lambda,\cdot)\in L^2(I_\pm)$, $\varphi(\lambda,\cdot) \in L^2(I)$.
Then
\begin{equation}\label{eq:A_abC}
(\mathcal{A}\varphi-\lambda\varphi)=g,
\end{equation}
\begin{equation}\label{eq:Def+-C}
\mathcal{A}\psi_--\lambda\psi_-=0,\quad
\mathcal{A}\psi_+-\lambda\psi_+=0,\quad
\end{equation}
It follows from \eqref{eq:Def+-C} that $\psi_\pm(\lambda,\cdot)\in \mathfrak{N}_\lambda(A_\pm)$ and hence, by~\eqref{eTWeyl},
\begin{align}\label{eq:Def_pmC}
\psi_+'(\lambda,a_+)&=m_{+}(\lambda)\psi_+(\lambda,a_+),\\\label{TashkentCity}
\psi_-'(\lambda,a_-)&=-m_{-}(\lambda)\psi_-(\lambda,a_-).
\end{align}
Since $\psi$ and $\psi'$ are continuous, we get 
\begin{align}\label{eq:BC1C}
     \varphi(\lambda,a_-)=\psi_-(\lambda,a_-),\quad
     \varphi(\lambda,a_+)=\psi_+(\lambda,a_+),
\\
     \label{eq:BC3C}
     \varphi'(\lambda,a_-)=\psi_-'(\lambda,a_-),\quad
     \varphi'(\lambda,a_+)=\psi_+'(\lambda,a_+).
\end{align}

Therefore, it follows from \eqref{eq:Def+-C} and \eqref{eq:Def_pmC}--\eqref{eq:BC3C} that $\varphi$ is a solution of \eqref{eq:A_abC}, such that
\[
\begin{split}
    \varphi'(\lambda,a_-)=\psi_-'(\lambda,a_-)
    &=-m_{-}(\lambda)\psi_-(\lambda,a_-)\\
    &=-m_{-}(\lambda)\varphi(\lambda,a_-),
\end{split}
\]
and
\[
\begin{split}
    \varphi'(\lambda,a_+)=\psi_+'(\lambda,a_+)
    &=m_{+}(\lambda)\psi_+(\lambda,a_+)\\
    &=m_{+}(\lambda)\varphi(\lambda,a_+).
\end{split}
\]
This proves (1) in Definition~\ref{def:4.1}.
\bigskip

\noindent
\textbf{2.} 
Conversely, let $g\in L^2(I)$ and let $\varphi$ be a solution of the problem~\eqref{eq:Afg2_4}, \eqref{eq:TBC10A}, \eqref{eq:TBC10B}.
Consider functions $\psi_\pm(\lambda,\cdot)\in \mathfrak{N}_\lambda(A_\pm)$ such that the conditions in~\eqref{eq:BC1C} hold.
Then, by~\eqref{eq:Def_pmC}, \eqref{TashkentCity},
\eqref{eq:BC1C}, \eqref{eq:TBC10A}, and
\eqref{eq:TBC10B}
we get
\begin{equation}\label{eq:Afg2_4_2}
\psi_\pm'(\lambda,a_\pm)=\varphi'(\lambda,a_\pm),
\end{equation}
and hence the functions
\begin{equation}\label{eq:Afg2_4_3}
\psi(\lambda,\cdot)=\begin{pmatrix}
\psi_-(\lambda,\cdot) & \varphi(\lambda,\cdot) & \psi_+(\lambda,\cdot)
\end{pmatrix}
\end{equation}
and $\psi'(\lambda,\cdot)$ are continuous on $\mathbb{R}$ and $\psi(\lambda,\cdot)\in\ker{(H-\lambda I)}$. Therefore, $\psi$ is a solution of \eqref{eq:Afg.3_4}. This proves (2) in Definition~\ref{def:4.1}.
\bigskip

\noindent
\textbf{3.} 
To prove (3) in Definition~\ref{def:4.1} we show that every 
$\lambda\in\rho(H)$ 
is a regular point of the problem~\eqref{eq:Afg2_4}, \eqref{eq:TBC10A}, \eqref{eq:TBC10B}.
Therefore, let $\lambda\in\rho(H)$,
$g\in L^2(I)$, and let $\wt g$ be its trivial extension  to $\dR$. Then, by
$\lambda\in\rho(H)$, there exists a solution $\psi(\lambda,\cdot)$ of~\eqref{eq:Afg.3_4}. Then following the arguments in Step \textbf{1} of the proof,  there exists a solution  $\varphi$  of the problem~\eqref{eq:Afg2_4}, \eqref{eq:TBC10A}, \eqref{eq:TBC10B}.
It remains to show that $\varphi$ is unique. For this,
assume that $\varphi$ is a solution of the problem
\begin{equation}\label{eq:Afg2_4_0}
  (\cA- \lambda I)\varphi=0\quad \text{on}\quad I
\end{equation}
such that \eqref{eq:TBC10A}, \eqref{eq:TBC10B} hold. As in Step \textbf{2}, consider functions $\psi_\pm(\lambda,\cdot)\in \mathfrak{N}_\lambda(A_\pm)$ such that the conditions in~\eqref{eq:BC1C} hold.
Then, as it was shown in Step \textbf{2},
\eqref{eq:Afg2_4_2} holds and hence the function $\psi$ given by~\eqref{eq:Afg2_4_3} is in $\ker{(H-\lambda I)}$.
Since $\lambda\in\rho(H)$, we obtain $\psi\equiv0$ on $\dR$ and hence also $\varphi\equiv 0$ on $I$ and $\lambda$ is a regular point of~\eqref{eq:Afg2_4}, \eqref{eq:TBC10A}, \eqref{eq:TBC10B}.

For the converse statement, assume that  $\lambda$  is a regular point of the problem~\eqref{eq:Afg2_4}, \eqref{eq:TBC10A}, \eqref{eq:TBC10B}. Let $\psi$ be in $\ker{(H-\lambda I)}$, i.e.,
\begin{equation}\label{Frankfurt}
  (\cA- \lambda I)\psi=0\quad \text{on}\quad \dR.
\end{equation}
Then, with the same arguments as in Step~\textbf{1}, we see that the restriction $\varphi$ of $\psi$ to $I$ solves~\eqref{eq:Afg2_4_0} such that \eqref{eq:TBC10A}, \eqref{eq:TBC10B} hold. As $\lambda$ is a regular point of the problem~\eqref{eq:Afg2_4}, \eqref{eq:TBC10A}, \eqref{eq:TBC10B}, there exists a unique solution of~\eqref{eq:Afg2_4_0}, which implies
$$
\varphi = \psi|_I \equiv 0.
$$
As the function $\psi$ is a solution of the differential equation~\eqref{Frankfurt} which is zero on an finite interval, we have 
$\psi \equiv 0$, that is, $\lambda$ is not an
eigenvalue of the operator $H$. Next, observe that $\wh\rho(A_-)\cap\wh\rho(A_+)$ coincides with the set of regular type points of the closed symmetric operator
$$
A_- \oplus A_{\rm in}\oplus A_+,
$$
where $A_{\rm in}$ is the minimal operator generated by $\mathcal{A}$ 
on $I$. 
Its domain consists of all functions $\varphi \in L^{2}(I)$ such that $\varphi$ and $\varphi'$ are locally absolutely continuous on $I$, $\cA \varphi \in L^{2}(I)$  and $\varphi(a_\pm) = \varphi^\prime(a_\pm)=0$. 
As the domains of the operators $
A_- \oplus A_{\rm in}\oplus A_+,
$ and $H$ are the same, except for finitely many dimensions, we have
$$
\wh\rho(A_-)\cap\wh\rho(A_+) = \wh\rho(H).
$$
As $\lambda\in\wh\rho(A_-)\cap\wh\rho(A_+)$
but not an eigenvalue, $\lambda \in \rho(H)$ follows.
\bigskip

\noindent
\textbf{4. }
To prove the uniqueness let us assume that $\tau_-$, $\tau_+$ is a pair of Herglotz-Nevanlinna functions such that \eqref{eq:TBC10.2_4} and~\eqref{eq:TBC10.3_4} are transparent boundary conditions for $\mathcal{A}$ on $I$, let $\lambda\in\dC_-\cup\dC_+$ and let $\varphi$ be a solution of~\eqref{eq:A_abC}, \eqref{eq:TBC10.2_4}--\eqref{eq:TBC10.3_4}. Then, by Definition~\ref{def:4.1}, there exists a solution $(\psi, \lambda)$ for~\eqref{eq::1} such that $\varphi$ is the restriction of $\psi$ to $I$ and, the restrictions $\psi_\pm$ to $I_\pm$ satisfy \eqref{eq:Def+-C}, i.e., $\mathcal{A}\psi_\pm-\lambda\psi_\pm=0$. Then as was shown in Step 1, $\psi_\pm$ satisfy equalities \eqref{eq:Def_pmC} and \eqref{TashkentCity}, and hence $\varphi$ satisfies equalities \eqref{eq:TBC1.1} and~\eqref{eq:TBC2.1}. Therefore,
\[
\tau_\pm(\lambda)=m_\pm(\lambda)\quad\text{for all }\quad \lambda\in\wh\rho(A_-)\cap\wh\rho(A_+).
\]
\end{proof}
Utilizing the notion of eigenvalues of problems with energy dependent boundary conditions from Definition~\ref{def:3.2} we directly obtain the following corollary.

\begin{corollary}
The set of eigenvalues of the problem~\eqref{eq:Afg2_4}, \eqref{eq:TBC10A}, \eqref{eq:TBC10B} coincides with $\sigma_p(H)$.
\end{corollary}
\begin{corollary}\label{cor:3.2}
Let under the assumptions of Theorem~\ref{thm:3.1.Cont} $\lambda\in \rho(H)$, $g\in{L^2(I)}$ and let $\psi=(H-\lambda I)^{-1}\wt g$, where $\wt g=g\oplus 0_{I_-\cup I_+}$.
Then the restriction $\varphi$ of $\psi$ to $I$  is the unique solution of the problem
\begin{equation}\label{eq:Afg}
  (\cA- \lambda I)\varphi=g
\end{equation}
subject to the transparent boundary conditions~\eqref{eq:TBC10A}--\eqref{eq:TBC10B}, i.e., the
solution $\varphi$ is given with the help of the compressed resolvent of $H$
\[
\varphi =P_{L^2(I)}(H-\lambda I)^{-1}\wt g,
\]
where $P_{L^2(I)}$ stands for the orthogonal projection in $L^2(\mathbb R)$ onto $L^2(I)$.
\end{corollary}

\section{Examples}\label{Examples}

In this section we apply the obtained results to special cases of the harmonic potential and the P\"oschl-Teller potential.

\subsection{Harmonic potential}
We consider the spectral  problem for the quantum harmonic oscillator
\begin{equation}
\left(-\frac{d^2}{dx^2}+\frac14 x^2\right)\varphi={E}\varphi, \ \ \ x\in \mathbb{R} \label{eq::ho}
\end{equation}
and (H1) and (H2) from Section~\ref{Shymbulak}
are satisfied.
The eigenvalues of the self-adjoint operator $H$ 
in $L^2(\dR)$ generated by the differential expression $\mathcal{A}$ 
(cf.\ Section~\ref{Shymbulak}),
\begin{equation}
  \mathcal{A}=\left(-\frac{d^2}{dx^2}+\frac14 x^2\right), \label{eq:HarOsc}
\end{equation}
 are ${E}_n=n+\frac12$ $(n\in\dN_0)$ and the corresponding eigenfunctions are the Chebyshev-Hermite functions
\[
\varphi_n(x)=
e^{x^2/4}\frac{d^n}{dx^n}\left(e^{-x^2/2}\right),\quad n\in\dN_0.
\]
For arbitrary ${E}$ any solution of \eqref{eq::ho} can be represented as a linear combination of two Weber parabolic cylinder functions $U(-{E},x)$ and $V(-{E},x)$, ~\cite[Sect. 19.3]{AbrSt65}. The functions  $U(-{E},x)$ and $V(-{E},x)$ have the following asymptotic for large $x$
\[
U(-{E},x)=e^{-\frac14 x^2}x^{{E}-\frac12}\left(1+O\left(\frac{1}{x^2}\right)\right),\quad x\to\infty,
\]
\[
V(-{E},x)=\sqrt{\frac{2}{\pi}}e^{\frac14 x^2}x^{-{E}-\frac12}\left(1+O\left(\frac{1}{x^2}\right)\right),\quad x\to\infty,
\]
see~\cite[Sect. 19.8]{AbrSt65}, and thus $U(-{E},\cdot)\in L^2(a_+,\infty)$
for every $a_+\in\dR$. By~\eqref{eq:Weyl_Func}, the Titchmarsch-Weyl coefficient $m_+({E})$ of $\cA$ on $(a_+,\infty)$ takes the form
\begin{equation}
   m_+({E})=\frac{U'(-{E},a_+)}{U(-{E},a_+)}. \label{eq:m+}
\end{equation}

Alongside with $U(-{E},x)$ the function $\widetilde U(x):=U(-{E},-x)$ is also a  solution to \eqref{eq::ho}, and $\widetilde U\in L^2(-\infty,a_-)$ for every $a_-\in\dR$.
The Titchmarsch-Weyl coefficient $m_-({E})$ of $\cA$ on $(-\infty,a_-)$ has the form
\begin{equation*}
   m_-({E})=-\frac{\widetilde U'(a_-)}{\widetilde U(a_-)}=\frac{U'(-E,-a_-)}{U(-E,-a_-)}. \label{eq:m-}
\end{equation*}

Let us show that the eigenvalues of the problem on the finite interval $I=(a_-,a_+)$
\begin{equation}\label{Jambul}
  (\cA- E I)\varphi=0\quad \text{on}\quad I
\end{equation}
with transparent boundary conditions
(see Theorem~\ref{thm:3.1})
\eqref{eq:TBC1.1}-\eqref{eq:TBC2.1} coincides with the set $\{n+\frac12\}_{n=0}^\infty$. Let $\varphi$ be a nontrivial solution of~\eqref{Jambul}, \eqref{eq:TBC1.1}-\eqref{eq:TBC2.1}.
Since $\varphi$ satisfies~\eqref{eq:TBC2.1}, it is proportional to $U(-{E},x)$, i.e.,
\[
\varphi(x)=CU(-{E},x),\quad x\in(a_-,a_+)
\]
for some $C\in\dC$.
Next, let us recall the identities
\begin{equation}\label{eq:Upmx}
\begin{split}
    &\pi V(-{E},x)\\
    &=\Gamma(\frac12-{E})\left\{-\sin(\pi{E})U(-{E},x)+U(-{E},-x) \right\}
\end{split}
\end{equation}
and
\begin{equation}\label{eq:Wronski}
  W(U,V)
  =\sqrt{{2}/{\pi}},
\end{equation}
where $W(U,V)$ is the Wronskian of the functions  $U$ and $V$,
see~\cite[19.4.1-2]{AbrSt65}.
Then, by~\eqref{eq:TBC1.1}, \eqref{eq:Upmx} and~\eqref{eq:Wronski},
\[
\begin{split}
   0&=(\varphi'(a_-)+m_-(E)\varphi(a_-))U(-{E},-a_-)\\
    &=-C W(U,\wt U)\\
    &=-\frac{C\pi}{\Gamma(1/2-{E})}W(U,V)=\frac{-\sqrt{2\pi}}{\Gamma(1/2-{E})}C.
\end{split}
\]
This implies that $E=E_n=n+1/2$, $n\in\mathbb{N}_0$. Moreover, the eigenfunction $\varphi$ coincides with the restriction of $CU(-{E_n},x)$ to the interval $(a_-,a_+)$. Therefore, boundary  conditions \eqref{eq:TBC1.1}-\eqref{eq:TBC2.1} are transparent for the expression $\mathcal{A}$ on $(a_-,a_+)$.

\subsection{P\"oschl-Teller potential}

The Schr\"odinger equation with P\"oschl-Teller potential
$$
\left(-\frac{d^2}{dx^2}-\frac{\ell(\ell+1)}{\cosh^2x\,}\right)\varphi={\lambda}\varphi, \ \ \ x\in\mathbb{R},
$$
has the solution $\varphi(x)=P^\mu_\ell(\tanh{(x)})$, where $P^\mu_\ell$ are the Legendre functions, and the eigenvalues (energies) are $\lambda=-\mu^2$ for $\ell\in \mathbb N$ and $\mu=1,\ldots,\ell$.
It satisfies  (H1).
Let us consider the spectral  problem for $\ell=1$, i.e. 
\begin{align}
\left(-\frac{d^2}{dx^2}-\frac{2}{\text{cosh}^2x\,}\right)\varphi={\lambda}\varphi, \ \ \ x\in {I_\pm},\label{eq:Poeschl}
\end{align}
Since the potential $V(x)=-\frac{2}{\text{cosh}^2x\,}$ tends to 0 at $\infty$, the Hamiltonian $H$ has continuous spectrum  $\sigma_c(H)=[0,+\infty)$.
Moreover, it is easy to check that $H$ has one negative eigenvalue $E=-1$.
Let us find the transparent boundary conditions for the P\"oschl-Teller operator $\mathcal{A}$ on the interval $I=(a_-,a_+)$.

The equation~\eqref{eq:Poeschl} has two linearly independent solutions
\[
u_\pm(\lambda,x)=\cos (x-a_\pm)\sqrt{\lambda}-\text{tanh\,}x\,\frac{\sin (x-a_\pm)\sqrt{\lambda}}{\sqrt{\lambda}}
\]
and
\[
v_\pm(\lambda,x)=\sin (x-a_\pm)\sqrt{\lambda}+\text{tanh\,} x\,
\frac{\cos (x-a_\pm)\sqrt{\lambda}}{\sqrt{\lambda}}
\]
and neither of them belong to $L^2(I_\pm)$. 
However,
\[
\psi_\pm(\lambda,x):=u(x,\lambda)\pm iv(x,\lambda)
=e^{\pm i(x-a_\pm)\sqrt{\lambda}}\left(1\pm\frac{i\text{tanh\,}x\,}{\sqrt{\lambda}}\right)
\]
is a solutions of~\eqref{eq:Poeschl} with $\psi_\pm(\lambda,\cdot)\in L^2(I_\pm)$. Let $A_\pm$ be the \emph{minimal operators} associated in $L^{2}(I_\pm)$, cf.\ Section~\ref{Almaty}, with the differential expression~\eqref{eq:Poeschl}.
Taking boundary triples for the operators $A_\pm$ in the form~\eqref{eq:BTforSL} we get
\[
\Gamma_0^\pm \psi_\pm=\psi_\pm(\lambda,a_\pm)=1\pm\frac{i}{\sqrt{\lambda}}\text{tanh\,}a_\pm,
\]
\[
\Gamma_1^\pm \psi_\pm=\pm \psi'_\pm(\lambda,a_\pm)=\mp\text{tanh\,}a_\pm +\frac{i}{\sqrt{\lambda}}\left(\lambda +\frac{1}{\text{cosh}^2a_\pm}\right).
\]
By~\eqref{eq:Weyl_Func}, the corresponding Weyl functions of $A_\pm$ is
\[
  m_\pm(\lambda)=\frac{\Gamma_1^\pm \psi_\pm(\lambda,\cdot)}{\Gamma_0^\pm \psi_\pm(\lambda,\cdot)}=i\sqrt{\lambda}
  \pm\frac{1}{\text{cosh}^2a_\pm(\text{tanh\,}a_\pm\mp i\sqrt{\lambda})}.
\]
Therefore, by Theorem~\ref{thm:3.1.Cont}, the transparent boundary conditions for the P\"oschl-Teller operator $\mathcal{A}$ on the interval $I=(a_-,a_+)$ take the form
\[
\varphi'(\lambda,a_-)+\left(i\sqrt{\lambda}
-\frac{1}{\text{cosh}^2a_-(\text{tanh\,}a_-+i\sqrt{\lambda})}\right)\varphi(\lambda,a_-)=0,
\]
\[
\varphi'(\lambda,a_+)-\left(i\sqrt{\lambda}
+\frac{1}{\text{cosh}^2a_+(\text{tanh\,}a_+-i\sqrt{\lambda})}\right)\varphi(\lambda,a_+)=0.
\]

Now, by applying above transparent boundary conditions to the solution $\varphi(x)=P_\mu^\ell(\tanh(x))$, one can show that the eigenvalue for \eqref{eq:Poeschl} is equal $-1$. It turns out that for $\ell=1$
$$
\varphi(x)=P_1^1(\tanh(x))=-\sqrt{1-\tanh^2(x)},
$$
and for its derivative one has
$$
\varphi'(x)=\frac{\tanh{(x)}}{\cosh^2{(x)}\sqrt{1-\tanh^2(x)}}.
$$
By substituting these in the equation for the transparent boundary conditions of the left boundary ($x=a_-$) we get
$$
i\sqrt{\lambda}-\frac{1}{\cosh^2{(a_-)}(\tanh{(a_-)}+i\sqrt{\lambda})}=\\
\tanh{(a_-)}.
$$
This leads to
$$
\lambda=-\left(\tanh^2{(a_-)}+\frac{1}{\cosh^2{(a_-)}}\right)=-1.
$$
Similarly, it can be shown for $a_+$.

%
%

\section{Conclusions}

In this work, we have presented a general approach for deriving transparent boundary conditions (TBCs) for the stationary Schrödinger equation with arbitrary potentials. By using the Weyl-Titchmarsh theory, we demonstrated that the TBCs can be expressed in terms of the Weyl-Titchmarsh coefficients, providing a unified and systematic method for treating a wide range of potentials. 

To validate the effectiveness of our method, we applied it to two specific cases: harmonic  and  Pöschl-Teller potentials. These examples highlight the versatility and robustness of the proposed framework, confirming its utility for both well-known and more complex potential forms. The harmonic potential serves as a classic example where our method recovers known results, while the Pöschl-Teller potential illustrates how the approach can handle potentials with more intricate structures.

Overall, this study provides a comprehensive and flexible toolkit for the derivation of TBCs in quantum mechanics, paving the way for future explorations of more complex systems. The ability to express TBCs through Weyl-Titchmarsh coefficients opens new avenues for analyzing quantum systems with arbitrary potentials, with potential applications in fields such as quantum optics, condensed matter physics, and nanotechnology. Future research could extend this framework to other evolution equations (e.g. Dirac equation) and for other domains like 2D and branched domains, broadening the applicability of our results in various quantum systems. The results obtained in this paper can be used for modelling of ballistic quantum transport in quantum materials and tunable band spectra in low-dimensional structures.

\section*{Appendix. Transparent boundary conditions and the coupling approach}

In this appendix we will embed  Sections~\ref{Shymbulak} and
\ref{Astana} in a more general setting. For this, we use results from the well-studied coupling method for two arbitrary symmetric operators \cite{DHMS}. This allows to view the Hamiltonian $H$ from Sections~\ref{Shymbulak} and \ref{Astana} as the coupling of two symmetric operators, one corresponds
to a finite interval (``the interior'') and 
the other one corresponds to the complement (``the exterior''). 

We recall the definition of boundary triples of a closed densely defined 
symmetric operator that has been introduced by Kochubei~\cite{Koc75}
(see also~\cite{GG} and references therein).

\begin{definition}\label{een1}
Let $S$ be a closed densely defined symmetric operator in a Hilbert space $\sH$.
A collection $(\cH, \Gamma_0, \Gamma_1)$ consisting of an auxiliary Hilbert space $\cH$ and two linear mappings $\Gamma_0$ and $\Gamma_1$ from $\dom S^*$ to $\cH$, is said to be a boundary triple for $S^*$ if:
\begin{enumerate} \def\labelenumi{\rm (\roman{enumi})}
\item[(a)] For all $f, g\in D(S^*)$ the identity holds
\begin{equation}\label{Greendef1}
(S^*f, g)-(f, S^*g) = (\Gamma_1 f, \Gamma_0 g)_{\cH}- (\Gamma_0 f, \Gamma_1g)_{\cH}.
\end{equation}

\item[(b)] The mapping
$\Gamma:=\begin{pmatrix}
         \Gamma_0 \\
         \Gamma_1
         \end{pmatrix}:\,D(S^*) \to {\cH}^2$ is surjective.
\end{enumerate}
\end{definition}

We emphasize that the  equality $n_+(S)=n_-(S)$ is a necessary and sufficient condition for the existence of a boundary triple for $S^*$.
Actually, this number corresponds to the dimension of the auxiliary Hilbert space $\cH$ in Definition~\ref{een1}.

Boundary triple provides a very efficient way to describe all self-adjoint extensions of
$S$. In particular, the operator $S_0:=S^*\upharpoonright_{\ker \Gamma_0}$ is a self-adjoint extension of $S$.

For every point $\lambda\in\dC_+\cup\dC_-$
the formula
\begin{equation}\label{GWeyl}
 M({\lambda})\Gamma_0 f_{\lambda}=\Gamma_1f_{\lambda},\quad
 f_{\lambda}\in\sN_{\lambda},
\end{equation}
correctly defines the Weyl function $M$ of $S$ corresponding to the
boundary triple $(\cH,\Gamma_0,\Gamma_1)$, see~\cite{DM91}.
The Weyl function $M$ is a Herglotz-Nevanlinna function.

An example of a boundary triple with corresponding Weyl function is given by the triples $(\dC,\Gamma_0^{\pm},\Gamma_1^{\pm})$ for $A_\pm^*$  in Section~\ref{Almaty}, see \eqref{eq:BTforSL}. Here, equation~\eqref{Greendef1}
follows from~\eqref{Greendef}. The corresponding
Weyl function satisfies \eqref{eq:Weyl_Func}, which is an example for \eqref{GWeyl}.

We recall the following coupling theorem~\cite{DHMS}.

\begin{theorem}
\label{thm:Coupl2}
Let $A_{\rm in}$ and $A_{\rm ex}$ be two
closed densely defined symmetric operators in the Hilbert spaces $\sH^{\rm in}$ and $\sH^{\rm ex}$, respectively, such that
$$
n_+(A_{\rm in})=n_-(A_{\rm in})=n_+(A_{\rm ex})=n_-(A_{\rm ex})=n.
$$
Let $(\cH,\Gamma_0^{\rm in},\Gamma_1^{\rm in})$ and
$(\cH,\Gamma_0^{\rm ex},\Gamma_1^{\rm ex})$
be boundary triples for   $A_{\rm in}^*$ and $A_{\rm ex}^*$, and let
$M^{\rm in}$ and 
$M^{\rm ex}$ 
be the corresponding  Weyl functions 
of $A_{\rm in}$ and $A_{\rm ex}$, respectively.
Set $A=A_{\rm in}^* \oplus A_{\rm ex}^*\upharpoonright_{\dom A} $ with
\begin{equation}
\label{coupl2}
 \dom A\!=\!\left\{\begin{pmatrix}
                f_1 \\
                f_2
              \end{pmatrix}\!\in\!
              \begin{pmatrix}
                \dom A_{\rm in}^* \\
                \dom A_{\rm ex}^*
              \end{pmatrix}\!:
\begin{array}{c}
  \Gamma_0^{\rm in} f_1-\Gamma_0^{\rm ex} f_2=0 \\
  \Gamma_1^{\rm in} f_1+\Gamma_1^{\rm ex}f_2=0
\end{array}
\right\}
\end{equation}
and let $S$ be the restriction of $A$ to the domain
\begin{equation}
\label{dom S}
  \dom S=\left\{\,f=\begin{pmatrix}
  f_1 & f_2  
  \end{pmatrix}^T\in \dom A:
  \Gamma_0^{\rm in} f_1=0 \,\right\}.
\end{equation}
Then:
\begin{enumerate}
\def\labelenumi{\rm (\roman{enumi})}
\item $S$ is a closed densely defined 
 symmetric operator with defect numbers $(n,n)$ and $S^*=A_{\rm in}^* \oplus
A_{\rm ex}^*\upharpoonright_{D( S^*)} $, where
\begin{equation}
\label{coupl2*}
\dom S^*=\left\{\begin{pmatrix}
                f_1 \\
                f_2
              \end{pmatrix}\in
              \begin{pmatrix}
                \dom A_{\rm in}^* \\
                \dom A_{\rm ex}^*
              \end{pmatrix}:
   \Gamma_0^{\rm in} f_1=\Gamma_0^{\rm ex}f_2
\right\}.
\end{equation}
\item $A$ is a self-adjoint extension of $S$.
\item
The compressed resolvent $P_{\sH^{\rm in}}( A-\lambda I)^{-1}\upharpoonright_{\sH^{\rm in}}$ of $S$ can be characterized by the following equivalence: For each $g\in\sH^{\rm in}$ and $\lambda\in\rho(A)\cap\wh\rho(A_{\rm ex})$ we have
    $f=P_{\sH^{\rm in}}( A-\lambda I)^{-1}(g\oplus 0)$ 
    if and only if $f$ is a solution of the ``boundary problem''
\begin{equation}\label{eq:ResINterBC}
  ( A_{\rm in}^*-\lambda I)f=g,\quad f\in D(A_{\rm in}^*) ,
  \end{equation}
with the energy dependent boundary condition
\begin{equation}\label{eq:ResINterBC2}
    \Gamma_1^{\rm in}f+M_{\rm ex}(\lambda)\Gamma_0^{\rm in}f=0.
\end{equation}
\end{enumerate}
\end{theorem}
The operator $A$ defined by formula~\eqref{coupl2} is called the \textit{coupling} of the operators $A_{\rm in}$ and $A_{\rm ex}$.
\begin{proof}
  The verification of~(i)--(ii) is straightforward, see for details~\cite[Proposition~4.3]{DHMS}.
  
  The compressed resolvent $P_{\sH^{\rm in}}( A-\lambda I)^{-1}\upharpoonright_{\sH^{\rm in}}$ of $S$ was calculated in~\cite[Theorem 5.9]{DHMS}, its characterization in the form (iii) was presented in~\cite[Proposition~5.11]{DHMS}, see also \cite[Section 2]{DM91}.
\end{proof}

Let us apply Theorem~\ref{thm:Coupl2} to the 
situation from the Sections~\ref{Shymbulak} and~\ref{Astana}. Let again $\mathcal{A}$
stand for the  one-dimensional stationary Schr\"{o}dinger operation which satisfies (H1) given by
$$
\mathcal{A}=-{d^2}/{dx^2}+V
\quad \mbox{on } \mathbb R
$$
with a  potential $V\in L^1_{loc}(\mathbb{R})$. 
Again, the operator $H$ generated by the differential expression $\mathcal{A}$ is self-adjoint in $L^2(\dR)$, cf.\ Section~\ref{Astana}.
As in Sections~\ref{Almaty}--\ref{Astana}
we consider the intervals
$$
I_-=(-\infty,a_-], \quad
I=[a_-,a_+], \quad \mbox{and} \quad
I_+=[a_+,\infty).
$$
For obvious reasons,  $I_-\cup I_+$ is considered to be the exterior and $I$ the interior. For the interior and the exterior we will define symmetric operators.

The operator ${A_{\rm ex}}$ corresponding to the exterior $I_-\cup I_+$ is defined as the orthogonal sum  of the operators $A_-$ and $A_+$ from Section~\ref{Almaty},
$$
A_{\rm ex}:=A_-\oplus A_+.
$$
As mentioned above, the triples $(\dC,\Gamma_0^{\pm},\Gamma_1^{\pm})$ 
are boundary triples for $A_\pm^*$,  see \eqref{eq:BTforSL}.
The corresponding Weyl functions $m_\pm$
 satisfy \eqref{eq:Weyl_Func}.
The orthogonal sum of the two boundary triples $(\dC,\Gamma_0^{\pm},\Gamma_1^{\pm})$  is a boundary triple for $A_{\rm ex}^*$. More precisely, a boundary triple for $A_{\rm ex}^*$ is defined on
$\psi_{\rm ex}=\begin{pmatrix}
                \psi_- & \psi_+
               \end{pmatrix}^T\in \dom A_-^*\oplus \dom A_+^*$
by
\begin{equation}\label{eq:BT_pm}
\Gamma_0^{\rm ex}\psi_{\rm ex}=\begin{pmatrix}
                        \psi_-(a_-) \\
                        \psi_+(a_+)
                      \end{pmatrix},\quad
\Gamma_1^{\rm ex}\psi_{\rm ex}=\begin{pmatrix}
                        -\psi_-'(a_-) \\
                        \psi_+'(a_+)
                      \end{pmatrix}.
\end{equation}
The corresponding Weyl function $M_{\rm ex}$ takes the form
\begin{equation}\label{eq:M_2}
  M_{\rm ex}(\lambda)=\begin{pmatrix}
                     m_-(\lambda) & 0 \\
                     0 & m_+(\lambda)
                   \end{pmatrix}.
\end{equation}

Let $A_{\rm in}$ be the minimal operator generated by $\mathcal{A}$ 
in the interior domain $I$. 
Its domain  consists of all functions   $\varphi \in L^{2}(I)$ such that $\varphi$ and $\varphi'$ are locally absolutely continuous on $I$, $\cA \varphi  \in L^{2}(I)$  and $\varphi(a_\pm) = \varphi^\prime(a_\pm)=0$. The
operator $A_{\rm in}$ is symmetric with deficiency indices $(2,2)$. The domain of the adjoint operator $A_{\rm in}^*$ consists of functions $\varphi\in L^{2}(I)$ such that $\varphi$ and $\varphi'$ are locally absolutely continuous on $I$ and  $\cA \varphi\in L^{2}(I_\pm)$. A boundary triple for $A_{\rm in}^*$ can be chosen as  $(\dC^2,\Gamma_0^{\rm in},\Gamma_1^{\rm in})$, see~\cite[Section 9]{DM91},
for $\varphi\in \dom A_{\rm in}^*$
\begin{equation}\label{eq:BT_1}
\Gamma_0^{\rm in}\varphi=\begin{pmatrix}
                        \varphi(a_-) \\
                        \varphi(a_+)
                      \end{pmatrix},\quad
\Gamma_1^{\rm in}\varphi=\begin{pmatrix}
                        \varphi'(a_-) \\
                        -\varphi'(a_+)
                      \end{pmatrix}.
\end{equation}

Let $A$ defined by formula~\eqref{coupl2} be the coupling of the operators $A_{\rm in}$ and $A_{\rm ex}$.
Then it follows from~\eqref{coupl2}, \eqref{eq:BT_pm} and~\eqref{eq:BT_1} that any function $\psi \in \dom A$,
\[
\psi=
\left\{\begin{array}{ll}
        \psi_- & \text{on }I_-; \\
        \varphi & \text{on }I; \\
        \psi_+ & \text{on }I_+,
\end{array}\right.
\]
satisfies the conditions
 \begin{align}\label{eq:BC1A}
        \varphi(a_-)=\psi_-(a_-),\quad
        \varphi'(a_-)=\varphi'_-(a_-),\\
        \label{eq:BC3A}
        \varphi(a_+)=\psi_+(a_+),\quad
        \varphi'(a_+)=\psi'_+(a_+),
\end{align}
and, hence, $\psi$ belongs to $\dom H$. The converse inclusion is also clear and we obtain
$$
H=A.
$$

Let $\lambda\in \rho(H)$ and  $g\in{L^2(I)}$. Observe that the energy dependent boundary condition~\eqref{eq:ResINterBC2} coincide
with the transparent boundary conditions~\eqref{eq:TBC10A}--\eqref{eq:TBC10B}.
Now the statement (iii) of Theorem~\ref{thm:Coupl2}  proves that every solution $f\in L^2(I)$ of the problem
\begin{equation}\label{eq:Afg_A}
  (\cA- \lambda I)f=g
\end{equation}
with  transparent boundary conditions~\eqref{eq:TBC10A}--\eqref{eq:TBC10B} coincides with the restriction of $(H-\lambda I)^{-1}\wt g$, where $\wt g=g\oplus 0_{I_-\cup I_+}$,
to the interval $I$, see also
 Corollary~\ref{cor:3.2}.

\begin{remark}\label{rem:4Halfline}
For a self-adjoint operator $H$ generated by the differential expression $\cA$ on the half-line $\dR_+=(0,+\infty)$ and the Dirichlet boundary condition at $0$
\begin{equation}\label{eq:TBC10.2}
    \varphi(\lambda,0)=0,
\end{equation}
the boundary condition
\begin{equation}\label{eq:TBC10.3}
      \varphi'(\lambda,a_+)+\tau_+(\lambda)\varphi(\lambda,a_+)=0,
\end{equation} 
is called transparent for $\cA$ on the interval $I=(0,a_+)$, if  the problem \eqref{eq:Afg_A}, \eqref{eq:TBC10.2}, \eqref{eq:TBC10.3} 
satisfies conditions (1)--(3) of Definition~\ref{def:4.1}.

Application of Theorem~\ref{thm:Coupl2} to the minimal operators $A_{\rm in}$ and $A_{\rm ex}=A_{+}$ generated by $\cA$ in $L^2(I)$ and $L^2(a_+,\infty)$, respectively,  shows that the transparent boundary condition for $\cA$ at point $a_+$ has the form~\eqref{eq:TBC10B}.
\end{remark}




\begin{thebibliography}{0}
\bibitem{Engquist1} B. Engquist,  A. Majda, Math. Comput. {\bf 81},  629 (1977). 
\bibitem{Engquist2} B. Engquist,  A. Majda, Commun. Pure Appl. Math. {\bf 32} , 313 (1979).
\bibitem{Gander} M. J. Gander,  L. Halpern, Mathematics of Computations, \textbf{74}, 153 (2004).
\bibitem{Antoine1} X. Antoine, E. Lorin, Q. Tang, Mol. Phys., \textbf{115}, 1861 (2017).
\bibitem{Arnold1998} A. Arnold and M. Ehrhardt, J. Comput. Phys., \textbf{145(2)}, 611 (1998).
\bibitem{Ehrhardt1999} M. Ehrhardt, VLSI Design, \textbf{9(4)}, 325 (1999).
\bibitem{Ehrhardt2001} M. Ehrhardt and A. Arnold, Riv. di Math. Univ. di Parma, \textbf{6(4)}, 57 (2001).
\bibitem{Hammer2014} R. Hammer, W. P\"otz, A. Arnold, J. Comput. Phys., \textbf{256}, 728 (2014).
\bibitem{Wu} X. Wu, J. Zhang, J. Comput. Math., \textbf{29}, 74 (2011).
\bibitem{Schwendt} M. Schwendt, W. P\"otz, Comput. Phys. Commun., \textbf{229}, 129  (2018).
\bibitem{Han0} Z. Xu, H. Han, Phys. Rev. E, \textbf{74}, 037704 (2006).
\bibitem{Antoine} X. Antoine, Ch. Besse, and S. Descombes, SIAM J. Numer. Anal., \textbf{43}, 2272 (2006).
\bibitem{Matthias2008} A. Zisowsky and M. Ehrhardt, Math. and Comput. Modell., \textbf{47}, 1264 (2008).
\bibitem{Zhang} J. Zhang, Z. Xu, X. Wu, Phys. Rev. E, \textbf{79}, 046711 (2009).
\bibitem{Han} H. Han, Z. Zhang, Appl. Num. Math. \textbf{59}, 1568 (2009).
\bibitem{Li} H. Li, X. Wu, J. Zhang, Phys. Rev. E, \textbf{84}, 036707 (2011).
\bibitem{Zheng07} C. Zheng, SIAM J. Sci. Comput., \textbf{29(6)}, 2494 (2007).
\bibitem{Jambul} J. R. Yusupov, K. K. Sabirov, M. Ehrhardt, and D. U. Matrasulov, Phys. Lett. A {\bf 383} 2382 (2019).
\bibitem{Klein2011} P.\ Klein, X.\ Antoine, C.\ Besse and M.\ Ehrhardt, Commun.\ Comput.\ Phys., 10 (5), pp.\ 1280-1304 (2011).

\bibitem{Jambul2} J.R. Yusupov, K.K. Sabirov, M. Ehrhardt, and D.U. Matrasulov, Phys. Rev. E {\bf 100}(3-1) 032204 (2019).


\bibitem {Tit62}
    E.C.\ Titchmarsh,
    Eigenfunction Expansions, Part I,
    Clarendon Press,
    Oxford, 1962,
    204~pp.
\bibitem{DM91}
V.A.~Derkach and M.M.~Malamud,
Generalized resolvents and the boundary value problems for
hermitian operators with gaps,
J.\ Funct.\ Anal., 95 (1991), 1--95.

\bibitem{DHMS}
V.A. Derkach, S. Hassi, M.M. Malamud, and H.S.V. de Snoo,
Generalized resolvents of symmetric operators and admissibility,
Methods Funct.\ Anal.\ Topol., 6 (2000), 24--55.

\bibitem{GG}
V.~I.~Gorbachuk and M.~L.~Gorbachuk,
\textit{Boundary problems for differential operator equations},
Naukova Dumka, Kiev, 1984 (Russian).

\bibitem{Givoli91}
D.\ Givoli, Non-reflecting boundary conditions, J.\ Comput.\ Phys., 94 (1991), 1-29.  

\bibitem{Tsynkov98}
S.V.\ Tsynkov, Numerical solution of problems on unbounded domains, Appl. Numer. Math., 27 (1998), 465-532.

\bibitem{Atk81}
F.V.\ Atkinson, On the location of the Weyl circles, Proc.\ Roy.\ Soc.\ Edinburgh Section A, 88 (1981),
345-356.

\bibitem{Moyer2006} C.A.\ Moyer, Computing in Science \& Engineering, 8 (4), pp.\ 32-40, 2006.

\bibitem{Klein2010} P.\ Klein, X.\ Antoine, C.\ Besse and M.\ Ehrhardt, Progress in Industrial Mathematics at ECMI 2010. Mathematics in Industry, 17, Springer, Berlin, Heidelberg, 2012.

\bibitem{EhrZhe09}
M.\ Ehrhardt and C.\ Zheng,
Implementing exact absorbing boundary condition for
the linear one-dimensional Schrödinger problem with
variable potential by Titchmarsh-Weyl theory, Preprint.

\bibitem{Zhe07}
Ch.\ Zheng, Approximation, stability and fast evaluation of exact artificial boundary condition for the one-dimensional heat equation,
J.\ Comput.\ Math., 25 (2007), no.\ 6, 730–745.

\bibitem{Har84}
B.J.\ Harris, The asymptotic form of the Titchmarsh-Weyl $m$-function,
J.\ London Math.\ Soc., 30 (2), no.\ 1, 110–118, 1984.

\bibitem{Str66}
A.V.\ \v{S}traus,
One-parameter families of extensions of a symmetric operator,
Izv.\ Akad.\ Nauk SSSR Ser.\ Mat.,  30, 1325–1352, 1966.
    \bibitem{Koc75}
A.N.\ Kochubei,
\newblock On extentions of symmetric operators and symmetric binary relations,
\newblock  Matem. Zametki, 17(1), 41--48, 1975.

\bibitem{Naj69}
M.A.\ Naimark,
\newblock {Linear Differential Operators},
\newblock Nauka, Moscow, 1969.

\bibitem{Jambul1} J.R.~Yusupov, Kh.Sh.~Matyokubov, and K.K.~Sabirov, Nanosystems: phys., chem., math.,  11 (2), 145–152, 2020.

\bibitem{AbrSt65} M.\ Abramowitz and A.\ Stegun (Eds.), Handbook of Mathematical Functions, Dover, New York, 1965.

\end{thebibliography}
\end{document}